\documentstyle[amsfonts,aps,twocolumn]{revtex}
\input epsf
\begin{document}
\title{Two-species magneto-optical trap with $^{40}$K and $^{87}$Rb}
\author{J. Goldwin, S. B. Papp, B. DeMarco\cite{DeMarco}, and D. S. Jin\cite{Jin}}
\address{JILA, National Institute of Standards and Technology
and University of Colorado, Boulder, Colorado 80309}
\date{22 October 2001}
\maketitle

\begin{abstract}
We trap and cool a gas composed of $^{40}$K and $^{87}$Rb, using a
two-species magneto-optical trap (MOT). This trap represents the
first step towards cooling the Bose-Fermi mixture to quantum
degeneracy. Laser light for the MOT is derived from laser diodes
and amplified with a single high power semiconductor amplifier
chip. The four-color laser system is described, and the
single-species and two-species MOTs are characterized. Atom
numbers of $1\times 10^7$ $^{40}$K and $2\times 10^9$ $^{87}$Rb
are trapped in the two-species MOT. Observation of trap loss due
to collisions between species is presented and future prospects
for the experiment are discussed.
\\~\\PACS number(s): 32.80.Pj, 03.75.Fi, 05.30.Fk\\~\\
\end{abstract}

The first experimental realizations of Bose-Einstein condensation
in dilute atomic gases \cite{BEC1,BEC2,BEC3} brought with them an
ever-increasing interest in the quantum behavior of such systems.
These systems exhibit weak and controllable interactions, and are
typically simpler to describe theoretically than their condensed
matter counterparts. The quantum statistics of fermions, however,
initially prevented the production of a quantum degenerate Fermi
gas of atoms. Specifically, the challenge came in maintaining the
rethermalizing collisions necessary for forced evaporative cooling
of the gas --- the Pauli exclusion principle forbids $s$-wave
collisions between identical fermions at the ultralow temperatures
necessary to reach quantum degeneracy. To circumvent this
limitation, the first experiment to produce a quantum degenerate
Fermi gas \cite{debbiescience} used two spin states of a single
fermionic isotope, thus allowing the rethermalizing collisions
necessary for evaporative cooling. Sympathetic cooling of
fermionic atoms to quantum degeneracy using a thermal bath of
bosonic atoms has more recently been demonstrated in systems using
$^6$Li and $^7$Li \cite{Li1,Li2}.

In this paper we report on the simultaneous trapping of $^{40}$K
(a fermion) and $^{87}$Rb (a boson) using a two-species
magneto-optical trap (MOT).  This MOT will be used as a
pre-cooling stage prior to forced evaporation of the $^{87}$Rb and
sympathetic cooling of the $^{40}$K gas.  To produce the four
frequencies of light necessary to operate the MOT, we have
developed a relatively simple laser scheme that includes the use
of a single high power semiconductor amplifier.  With this system
we are able to trap $2\times 10^9$ $^{87}$Rb atoms and $1\times
10^7$ $^{40}$K atoms simultaneously. In addition we can monitor
either species during the operation of the MOT, and have observed
number loss in the $^{40}$K cloud due to the presence of trapped
rubidium.

\begin{figure}
\begin{center}\epsfxsize=3 truein \epsfbox{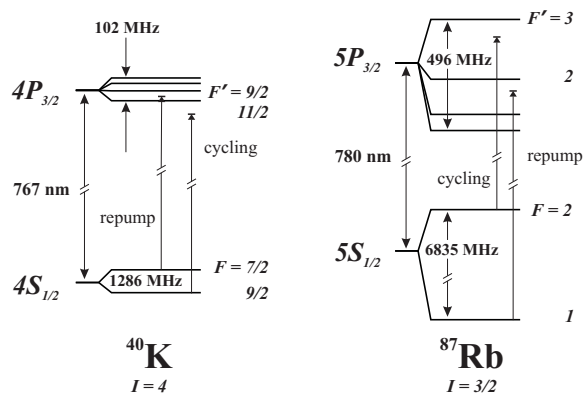}
\end{center}
\caption{Schematic hyperfine energy level diagrams for $^{40}$K
and $^{87}$Rb, showing the cycling and repump transitions used in
the experiment. Note the inverted structure and considerably
smaller splittings for $^{40}$K, whose excited state manifold
spans only $\sim 100$ MHz. $F$ and $F^\prime$ are the total atomic
angular momentum quantum numbers for the ground and excited
states, respectively, and $I$ is the nuclear spin quantum
number.\label{fig:energy_levels}}
\end{figure}

A MOT for trapping two different elements requires twice as many
laser frequencies as a single-species MOT. All of the light for
our MOT is generated by laser diodes and amplified by a single
high power tapered semiconductor amplifier chip (Toptica Photonics
TAE 780 \cite{disclaimer,PA}). The design of the laser system and
MOT optics exploits the similar wavelengths of the $D_2$ lines in
rubidium and potassium, whose energy levels are shown
schematically in Fig. \ref{fig:energy_levels}.

The generation of the laser light for the $^{87}$Rb MOT begins
with a grating-feedback stabilized external cavity diode laser
\cite{ECDL}.  The laser generates 20 mW of single mode,
narrow-band light. Some of this light is used to frequency lock
the laser via saturated absorption spectroscopy to the peak of the
$^{87}$Rb $F=2\to F^\prime = 2-3$ crossover line. The rest of the
light is then frequency shifted via a double-passed 80 MHz
acousto-optic modulator (AOM), and less than a milliwatt is sent
to a second laser diode for injection locking. The remaining light
is then available for optically pumping and/or probing the atom
sample.

The current to the injection locked laser is modulated to create
sidebands for hyperfine repumping
\cite{current1,current2,current3}. The modulation source is an
yttrium iron garnet crystal oscillator (Microsource Inc. MCO-0207)
whose output is coupled into the diode current using a bias ``T"
(Picosecond Pulse Labs 5585). In this way we produce repump and
cycling light for the MOT in a single beam and without the need
for a second external cavity diode laser. We can generate repump
light on either the $^{87}$Rb $F=1\to F^\prime = 1$ or $F=1\to
F^\prime = 2$ transition. Up to 10\% of the total 780 nm MOT light
is available for hyperfine repumping, which is sufficient for the
$^{87}$Rb MOT.

The 767 nm light for the $^{40}$K MOT is also generated using a
master-slave injection locking setup.  To reach the wavelength of
the potassium $D_2$ lines the laser diodes must be cooled using
multiple stages of thermo-electric coolers. In addition, a water
heat exchange plate was necessary to cool the slave laser to
$-40\,^\circ \mbox{C}$. Each laser is housed in a hermetically
sealed can with dessicant to prevent water condensation.

The 767 nm master laser is frequency locked to the peak of a
ground state feature of $^{41}$K via saturated absorption
spectroscopy \cite{K41note}.  The remaining light is then used to
injection lock the 767 nm slave laser diode for amplification.  A
$^{40}$K MOT typically requires more light for repumping than
$^{87}$Rb due to the small excited state hyperfine splitting
\cite{KMOT1,enrichedK} (see Fig. \ref{fig:energy_levels}). We
found a modulation scheme similar to that described above
incapable of stably generating sidebands with $>$15\% of the total
output power. To allow for more $^{40}$K repump power, we instead
frequency shift the light from the $\mbox{767 nm}$ slave laser via
a double-passed 500 MHz AOM for repumping. The unshifted light is
double-passed through a 110 MHz AOM to generate light on the
potassium cycling transition. The slave laser thus provides enough
power to generate both frequencies for the $^{40}$K MOT.

\begin{figure}
\begin{center}
\epsfxsize=3.5 truein \epsfbox{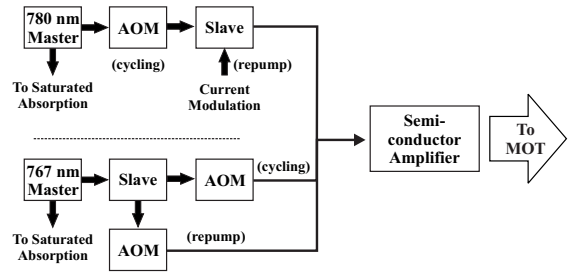}
\end{center}
\caption{Schematic of the laser system used in these experiments.
All four frequencies used for the two-species MOT are produced in
three independent beams and combined before injection into a
single semiconductor tapered amplifier chip.  The amplifier output
then consists of a single beam capable of trapping and cooling
$^{87}$Rb and $^{40}$K simultaneously.
 }\label{fig:lasers}
\end{figure}

High laser power for the MOT is obtained using a single tapered
semiconductor amplifier chip. A schematic summary of the laser
system is shown in Fig. \ref{fig:lasers}.  This type of
single-amplifier system has been used in experiments with two
isotopes of a single atom \cite{PA1,PA2} to provide amplification
with up to 12 GHz of bandwidth. In our system we exploit an
amplification bandwidth of $\mbox{7 THz}$. At $23\,^\circ
\mbox{C}$ the chip has an amplified spontaneous emission (ASE)
gain profile centered at 773 nm and with a full-width at
half-maximum (FWHM) of 16 nm. We measure an amplifier gain of
roughly 100 at 767 and 780 nm, and we can vary the amplified
powers in the four frequencies by controlling the relative powers
of the injected beams. This system produces all of the light
necessary for the two-species MOT in a single beam.

\begin{figure}
\begin{center}
\epsfxsize=3 truein \epsfbox{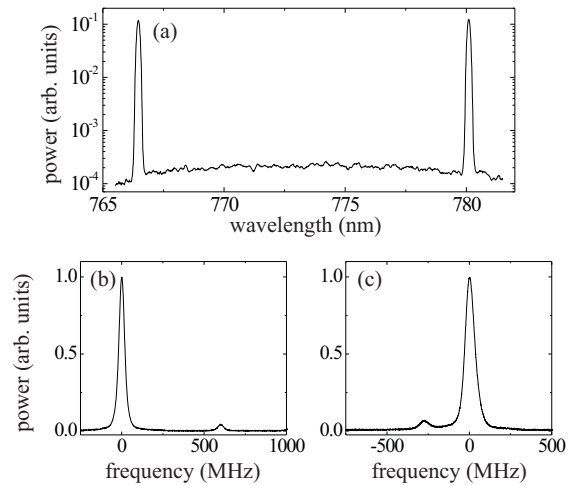}
\end{center}
\caption{Optical spectrum of the injected amplifier.  (a) The MOT
beam, measured with an optical spectrum analyzer, showing the
narrow 767 and 780 nm peaks above a highly suppressed ASE floor.
The widths of the peaks are resolution limited. (b) Closeup of the
$^{87}$Rb cycling and repump frequencies, measured with a
Fabry-Perot spectrometer with a free spectral range of 1.5 GHz.
The frequencies are referenced to the cycling transition. The
repump power shown here is approximately 4\% of the total 780 nm
light. (c) $^{40}$K trap and repump frequencies on the same
spectrometer. The repump power is approximately 6\% of the total
767 nm light. }\label{fig:amplifier_spectrum}
\end{figure}

The four frequencies injecting the amplifier are coupled together
with a series of polarizing beam splitter cubes and half-wave
retardation plates. Adjusting the orientations of the retarders
allows simple control of the relative power in each frequency sent
to the MOT.  Figure \ref{fig:amplifier_spectrum} shows the optical
spectrum of the injected amplifier. The single beam output from
the amplifier is expanded and shaped into a roughly gaussian beam
with a $1/e^2$ diameter of \mbox{3 cm}. Running the amplifier with
500 mW of total output light results in $\sim 300$ mW for the MOT
after beam shaping and spatial filtering. The beam from the
amplifier is then split three ways with each beam retroreflected
for the MOT.

The MOT itself is formed in a $2 \times 2 \times 6 \,\mbox{inch}$
rectangular glass cell, with a one inch diameter window at one end
for imaging. This cell is maintained at a pressure $\lesssim
10^{-10} \,\mbox{Torr}$. Alkali metal dispensers provide the
background vapor for the MOT.  The dispensers are housed in two
glass arms attached to the cell --- one for potassium and the
other for rubidium. The $^{87}$Rb source is a commercially
available dispenser (SAES Getters), while the potassium source is
made in-house from a KCl sample enriched in $^{40}$K (Trace
Sciences International), as described previously \cite{enrichedK}.
The vapor pressure for each species can be independently
controlled by varying the currents used to heat the dispensers.

In characterizing the behavior of the system we began by
optimizing each single-species MOT for number of trapped atoms. In
the case of $^{87}$Rb, with a peak light intensity of 70 mW/cm$^2$
at the MOT we are able to trap $2\times 10^9$ atoms in the absence
of trapped potassium. The atom number is determined from
fluorescence collected onto a photodiode or captured onto a charge
coupled device array (CCD). The $^{87}$Rb cloud is about 5\,-10 mm
in diameter. We find that a detuning of $\Delta\approx
-4\,\Gamma_{Rb}$ optimizes the number of trapped $^{87}$Rb atoms,
where $\Gamma_{Rb}=5.98 \,\mbox{MHz}$ is the natural linewidth of
the $D_2$ lines in rubidium.  As mentioned above we can repump on
either the $^{87}$Rb $F=1\to F^\prime = 1$ or $F^\prime = 2$
transition. We measured a slight increase in atom number when
repumping on the $F^\prime = 2$ transition. The magnetic field
gradient provided by our coils is typically 13-18 G/cm.

With no $^{87}$Rb MOT, and with 70 mW/cm$^2$ peak intensity of 767
nm light, we obtain a $^{40}$K cloud with $2\times 10^7$ atoms,
and measuring 0.1-0.3 mm in diameter.  The MOT is operated at a
detuning of $\Delta\approx -3\,\Gamma_K$, where $\Gamma_K=6.09
\,\mbox{MHz}$ is the natural linewidth of the $D_2$ lines in
potassium.  The low $^{40}$K number is due to the lower room
temperature vapor pressure of potassium. Our previous experience
with $^{40}$K suggests that more atoms can be trapped by heating
the glass cell, however we believe these numbers are sufficient
for our purposes. Because the $^{40}$K gas will be sympathetically
cooled in the next stage of the experiment, it is not necessary or
desirable to have a relatively large $^{40}$K MOT.

The two-species MOT is then operated with the same laser detunings
and magnetic field gradient used to optimize $N_{Rb}$ and $N_K$ in
the single-species MOTs.  With these parameters fixed, changing
the relative powers in each beam injecting the amplifier allows us
to tune the relative number $N_{Rb}/N_K$ from 100-500 while
maintaining at least $5\times 10^6$ trapped $^{40}$K atoms. This
will enable us to optimize the initial conditions for sympathetic
cooling. The $^{40}$K cloud, which is smaller in diameter, forms
completely within the center region of the larger $^{87}$Rb cloud.
Typical conditions of operation and the atom numbers obtained for
the single-species and two-species MOTs are compared in Table
\ref{tab:mots}.

We can independently monitor either species while the two-species
MOT operates. This permits in-situ measurement of $N_K$ and
$N_{Rb}$. We image fluorescence from the atoms onto a CCD array
and use narrow-band optical filters (CVI Laser Corp. F03-766.5-4,
F03-780.0-4) to selectively view either species.  These filters
have center (transmission) wavelengths of 767 and 780 nm,
respectively, and an optical depth of 4 outside the 3 nm FWHM
bandwidth.

\begin{figure}
\begin{center}
\epsfxsize=3 truein \epsfbox{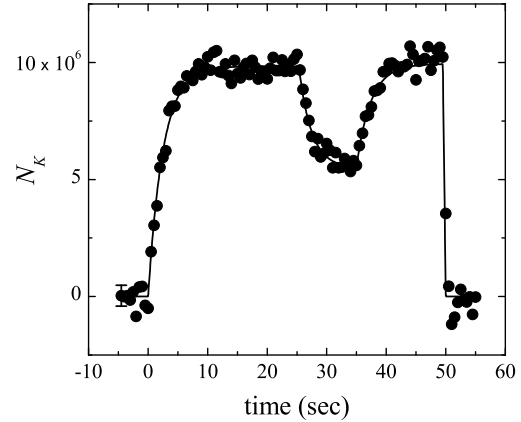}
\end{center}
\caption{Potassium MOT number loss in the presence of trapped
rubidium.  The data are averaged over nine consecutive
experiments, and the solid line is a fit to exponential growth and
decay in each region.  The period from 25 sec to 35 sec, where the
MOTs coexist, shows the loss in $^{40}$K atom number, $N_K$,
during operation of the two-species MOT.  For comparison, the
$^{40}$K loading rate is $4\times 10^6 \,\mbox{sec}^{-1}$, and the
time constant for exponential decay during the loss is
approximately 2 seconds. } \label{fig:number_loss}
\end{figure}

By taking a sequence of fluorescence images through the
appropriate filter, we observe the time evolution of either cloud
in the two-species MOT.  Figure \ref{fig:number_loss} shows the
results of an experiment monitoring the $^{40}$K cloud. Initially
all of the MOT light is on, but the 780 nm light is locked far off
resonance and the $B$-field is off so that no trapped atoms are
present. After turning on the field and letting the $^{40}$K MOT
evolve for 25 seconds, the 780 nm light is quickly shifted onto
the $^{87}$Rb cycling transition, allowing the rubidium MOT to
fill. After ten seconds in the presence of trapped $^{87}$Rb, the
780 nm light is jumped back off resonance and the $^{40}$K MOT is
allowed to evolve again freely.

In these data we observe a decrease in $N_K$ of 45\% in the
presence of the rubidium cloud.  This is attributed to
light-assisted heteronuclear collisions in the MOT.  A loss rate
of 20\% has been reported in a $^{85}$Rb\,--\,$^{39}$K system with
more evenly matched number between species \cite{bagnato}. If we
reverse the above experiment to monitor the $^{87}$Rb cloud for
loss in the presence of trapped $^{40}$K, we do not observe an
effect. This agrees with the observations reported by the Sao
Carlos group.

In summary we have demonstrated a two-species MOT for the
simultaneous trapping of $^{40}$K and $^{87}$Rb.  This trap will
serve as a pre-cooling stage prior to the sympathetic cooling of
the $^{40}$K to quantum degeneracy. Using a relatively simple
four-color diode laser scheme and a single high power
semiconductor amplifier, atom numbers of $N_{Rb}=2\times 10^9$ and
$N_K=1\times 10^7$ are obtained. In addition we have observed a
pronounced number loss in the $^{40}$K MOT due to heteronuclear
collisions in the presence of trapped $^{87}$Rb atoms. Current
work progresses on loading into a purely magnetic trap for
evaporative cooling.

\section*{Acknowledgements}
The authors would like to express their appreciation for useful
discussions with C. Wieman and E. Cornell.  In addition we would
like to acknowledge the work of H. Green, who built the glass
cells and enriched $^{40}$K sources used in these experiments. We
also acknowledge support from the U.S. Department of Energy,
Office of Basic Energy Sciences via the Chemical Sciences,
Geosciences and Biosciences Division.

\begin{table}\caption{Experimental parameters and atom numbers for the
single-species and two-species MOTs.   $I_c$ is the total peak
light intensity on the cycling transition and $I_r$ the total peak
intensity of hyperfine repumping light for the given
species.}\label{tab:mots}
\begin{tabular}{ccc}
 \rule[-1mm]{0mm}{4mm} & $^{40}$K & $^{87}$Rb \\
 \hline
 \rule[-3mm]{0mm}{7mm}$I_c$ (mW/cm$^2$) & 70 & 70 \\
 $I_r$ (mW/cm$^2$) & 4 & 2 \\
 $\Delta_c$ & $-3\Gamma_K$ & $-4\Gamma_{Rb}$ \\
 \hline
 \hline
 \rule[-3mm]{0mm}{7mm} $N$ (single-species) & $2\times 10^7$ & $2\times 10^9$\\
 $N$ (two-species) & $1\times 10^7$ & $2\times 10^9$ \\
\end{tabular}
\end{table}


\begin{thebibliography}{10}
\bibitem{PA}
    As a technical note for researchers interested in the lifetimes of
    amplifiers such as this, our chip has only been in use since March, 2001.
\bibitem{K41note}
    The $^{41}$K excited state manifold is only 17 MHz wide, so
    that its structure
    is not resolved by saturated absorption spectroscopy.  The
    entire $^{41}$K, $F=2\to F^\prime$ manifold therefore appears
    as a single spectroscopic feature; we frequency lock the laser
    to its peak.
\end{thebibliography}
\end{document}